\magnification 1200

to be published in {\it PHYSICA} {\bf A}
\vskip 1cm

\centerline {Monte Carlo Simulations of Sexual Reproduction} 

D.Stauffer*, P.M.C.de Oliveira, S.Moss de Oliveira, and R.M.Zorzenon dos Santos

Instituto de Fisica, Universidade Federal Fluminense,
av. Litoranea s/n, Boa Viagem, Niteroi RJ 24210-340, Brazil

*Present and permanent address: Institute for Theoretical Physics, Cologne 
University, D-50923 K\"oln, Germany

\bigskip
Abstract: Modifying the Redfield model 
of sexual reproduction and the Penna model of 
biological aging, we compare reproduction with and without recombination in
age-structured populations. In contrast to Redfield and in agreement with 
Bernardes we find sexual reproduction to be preferred to asexual one. In
particular, the presence of old but still reproducing males helps the survival 
of younger females beyond their reproductive age. 

\bigskip
\centerline {I. Introduction}

According to Genesis, work, sex, ageing, and death started simultaneously.
Most of today's higher species prefer sexual over asexual reproduction, and even
asexual species often employ some form of genetic recombination. There are
many theories but no consensus[1] why sex is preferred. For example, sex 
may spread advantageous mutations better through the population, sex increases
genetic variety which may help species to adjust after a drastic change in the
environment, or sex may protect against parasites. In a computer model with
only deleterious mutations, a constant environment, and no parasites, none of
these advantages hold. Indeed, in such a simulation 
Redfield[2] found only seldomly a clear 
advantage of sex to justify its cost to feed the males (which were also stated
[2] to be unreliable in transmitting the correct genes). Then, what are males
useful for ?

Charlesworth et al [3] have claimed that large sexual, as opposed to large
asexual, populations are protected against the mutational meltdown due to the
accumulation of deleterious inheritable mutations[4]. However, this claim was 
refuted [5] as being due to insufficient observation times: In their model
the decay time increases exponentially with population size. Different models
of Bernardes [5] with and without age structure gave clear advantages for sex:
The population can survive with sex better than without. More recently, however,
Bernardes [6] emphasized the advantages of asexual reproduction with genetic
recombination within the same individual (``meiotic parthenogenesis'') observed
in many asexual species. 

The present paper ignores these intermediate forms and compares sexual 
with asexual reproduction, assuming (as did Redfield[2]) genetic recombination
only in the sexual case. In contrast to Redfield we follow Bernardes[5,6] in 
assuming that only a fraction $h$ of the genetic mutations are dominant, the 
others being recessive. We want to see if now sex is more helpful in the   
Redfield model, using both her techniques (Chapter II) as well as Monte Carlo 
simulations based on a bit-string model of the genome (Chapter III). Then we 
use in Chapter IV the Penna bit-string
model [7,5] for age-structured populations to find out if the presence of males
allows females to live beyond their maximum reproductive age; for the asexual
case the accumulation of deleterious mutations kills all individuals beyond 
reproductive age[8] if parental care [9] is ignored. Our sexual Monte Carlo 
simulations are similar to those of ref.6 but include some 
features like male fidelity ignored by Bernardes.   

\bigskip
\centerline {II. Redfield Model}

The Redfield model [2] is a very fast algorithm avoiding all random
numbers and calculating from the probability distributions of the mutations
at time $t$ the corresponding distributions at the next time step $t+1$. It
should be exact in the limit of very large populations and cannot give the
finite-size effects studied in ref.6.  

The algorithm assumes from the beginning that the population is constant;
thus the birth rate must adjust to the temporal variation of the death rate. 
Thus the crucial effect of mutational meltdown due to the accumulation of only
deleterious inherited mutations [4] cannot be studied with this assumption. 
In principle this assumption is biologically unrealistic[5] though widespread
in the biological literature. It is known that many species have died out
even without any human intervention. Thus we have not used this dangerous
assumption in our Monte Carlo work of the following chapters. It seems, however,
that in the context of the Redfield calculations [2] this assumption is allowed.
Monte Carlo calculations[10] have confirmed the average survival probability
observed by Redfield in the asexual case and following from traditional 
steady-state theory. Moreover, we now have applied the Redfield method also to 
the case where mutational meltdown occurs, by assuming exactly one mutation 
per genome at each time step (``generation''). Then the mutational meltdown was
signalled clearly by an average survival probability decaying towards zero, 
whereas in ref.2 and in our corresponding simulations using a Poisson 
distribution of mutations these survival rates converged to a positive plateau
value. Thus the Redfield algorithm can signal mutational meltdown if it occurs,
and the assumption of a Poisson distribution can avoid this meltdown since now
with a nonzero probability no new mutation happens and the accumulation of an
unlimited number of mutations is avoided for some individuals. 

In her algorithm, Redfield starts from some progeny distribution $P(m), \;
m=0, 1, \dots$, giving the probability that an individual has $m$ genetic   
diseases in the genome. Darwinistic selection of the fittest then transforms
this $P(m)$ into a survivor distribution $L(m) \propto (1-s)^m P(m)$ giving
the probability that a survivor has $m$ deleterious mutations in the genome.
Here the selection coefficient $s$ was taken as 0.1, and the various mutations
(``genetic diseases'') are assumed to act independently (see below for the
alternative of truncation selection[2]). Now $n$ new hereditary mutations happen
according to a Poisson distribution $\mu^n \exp(-\mu)/n!$ so that the individual
has $m+n$ mutations; here $\mu$ of order unity is the mutation rate per genome
and generation. In the asexual case, the resulting distribution of mutations 
is already the new progeny distribution $P(m)$, and the above cycle of         
selection and mutation is repeated again and again. 

In the sexual case, according to Redfield [2]
the mutation rate for males can be larger by a factor 
$\alpha$ than that for females. Thus after  selection transformed the
progeny $P(m)$ distribution into the survivor distribution $L(m)$, mutations
of a rate $\mu$ produce the female distribution $F(m)$ and those of a rate
$\alpha \mu$ give the male distribution $M(m)$. Then male gametes (sperm cells)
are produced containing half of the male genome; thus their mutation
number $m_m$ is
roughly half of the number of mutations in the father's genome. Analogously,
the number $m_f$ of mutations in the female gametes (egg cells) is roughly
half of the mother's number of mutations. The fusion of two gametes adds 
these two numbers,
$$ m = m_m + m_f \quad ,\eqno (1)$$
to produce the mutations in the progeny distribution $P(m)$. Now the cycle
of selection, mutation, gamete production, and fusion is repeated  again
and again until the changes with each iteration become negligible. 

(More precisely, the mutations in each gamete are not exactly half of those
of the parent but follow a binomial distribution, which simulates the random
selection of the transmitted half of the genome via the processes of 
meiosis, crossover, and mitosis [5].)   

For efficiency purposes, the number of mutations was assumed to be
below $10^2$. Then 
this whole procedure gives within a few seconds on a workstation after less
than 100 iterations an average survival probability which barely changes 
with the number of iterations.  

In this model the fitness is the average survival probability
$$\sum_m P(m)(1-s)^m/\sum_m P(m) \quad ;$$ its steady state value is [2] 
exp$(-\mu)$ in the asexual and exp$(-\mu(1+\alpha)/2)$
in the sexual case, as confirmed by computer simulations. In particular,
for $\alpha = 1$ sex does not change at all the average fitness and thus fails
to justify its additional cost and complication.
However, it is obvious  from eq(1) that this Redfield model does not distinguish
between dominant and recessive mutations. Usually, only a rather small 
fraction $h$ of the
genetic diseases are dominant and reduce the survival probability even if only
one of the two parents (for sexual reproduction) carries the dangerous mutation;
the remaining mutations are recessive and affect the health only if present in 
both parents. Thus instead of eq(1) we could assume $m = (m_m + m_f)h$ due to 
dominant mutations. The recessive mutations are negligible if we have an 
infinite size of the genome where the finite number of mutations can be 
distributed; otherwise the recessive mutations produce $ r m_m m_f$ dangerous
diseases, where $r$ varies as the reciprocal size of the genome. Thus we now 
replace eq(1) by 
$$ m = (m_m + m_f)h + r m_m m_f \quad (h \ll 1, \, r \ll 1) . \eqno (2)$$
The original Redfield model has $h=1, \, r=0$ and from this point of view
corresponds to the limit of extreme inbreeding, which is known to be bad.
Nature usually has weeded out most of the dominant mutations, and therefore
most of the actually observed mutations are recessive. 

With eq(1) replaced by eq(2) in the algorithm, the advantages of sexual compared
with asexual reproduction become obvious even at the high value $\alpha = 10$
of male unreliability: For the Redfield value $\mu = 0.3$, asexual reproduction
gives a fitness of 0.74, whereas sexual reproduction gives a fitness of 0.95 to 
0.97 using eq(2) with $h=0.2$ and $0 \le r \le 0.2$, to be compared with 0.19 
using eq(1). Thus the cost of sex is more than justified by the drastic increase
of the survival probability, in contrast to its drastic decrease found by
Redfield [2]. Fig.1 gives an overview of the resulting fitness as a function of
parameters; cases where more than 1000 mutations became relevant are ignored.  

These computer simulations thus justify sex already from the fact that it
suppresses the effects of recessive mutations. 

Redfield[2] suggested instead to justify sex with the assumption that females
select only the youngest males for mating. Since there is no age-structure 
in this model, we instead follow Bernardes[5,6] by assuming that females
select only the healthiest males as mates. Then indeed similar drastic 
advantages of sex are found even if we use eq(1) without distinction between
recessive and dominant mutations. Thus only males with a number of mutations 
below a sexual truncation threshold $sextruc$ mutations
mate with females (independent of the female number of mutations). For 
$sextruc$=1
and 2 the fitness increased from 0.74 to 0.97 and 0.90, respectively. Fig.2 
shows some examples of the time evolution. 
(Following ref.2 we show here versus time first the simulation without sex, and 
then 
starting from the asexual equilibrium we show how for different values of the
sexual truncation parameter we get different results with fitness higher or    
lower than before.)  
For consistency we also replaced
the survival probability by a step function with a threshold similar to 
$sextruc$: 
Instead of $(1-s)^m$ the probability is 1 for up to 6 mutations, and 0
for more mutations, a truncation method already used by Redfield[2]. 

If there are as many males as females, then this method of selecting only
the healthiest males for mating means, that males mate with more than one
female. If instead we use monogamy in our probability distributions, then
this selection of males may diminish the average survival rate since many
females now do not find a suitably healthy mate.  

\bigskip
\centerline {III. Redfield-type Bitstring Model} 

While all mutations in the Redfield model (as well as those in this paper)
are hereditary, the method of probability distributions does not allow to
identify offspring with particular parents who share the same genetic diseases
(mutations). Thus an entirely different Monte Carlo simulation was made which
should reproduce qualitatively though not quantitatively the effects of the 
Redfield model in our previous chapter. For asexual reproduction, the genome
is then represented by one computer word containing 32 bits. Each bit is either
set (=1) or off (=0) and represents one serious inheritable disease. Thus 
perfect health corresponds to the whole computer word consisting of zeroes. 
If a word contains $m$ set bits, it represents $m$ dangerous mutations reducing,
as above, the survival probability to $(1-s)^m$ with $s = 0.1$. 

At each time step, each surviving individual gives birth to $b$ offspring,
and each offspring differs by $\mu$ mutations from the parent. Such a mutation
consists in selecting randomly one of the 32 bits, and setting this bit. If
the selected bit is already set it stays that way and nothing happens. In this
way, as with a Poisson distribution of mutations, with a nonzero probability
no new mutation is accumulated, and mutational meltdown[4] can be avoided. 
Therefore no complicated Poisson distribution was simulated, in contrast to 
ref.6; a value of $\mu < 1$ means that one such mutation was attempted with
probability $\mu$. In this version the simulation would give a population $N(t)$
which
asymptotically would decrease or increase exponentially with time $t$, depending
on the birth rate. To avoid the increase towards infinity we assume an 
additional death rate due to environmental restrictions of food and space. Then 
the survival probability is reduced by a Verhulst factor $1 - N(t)/N_{max}$
where the carrying capacity $N_{max}$ was taken as 200,000. 

For the sexual case, each individual has two such computer words as the genome.
Biological recombination of genes
is now a genetic algorithm: a crossing point between 0 and
32 is selected randomly to mix the genes of father and mother. The first
computer word of the child has the first bits up to this 
crossing point from the father, and the remaining bits 
from the mother. (We select randomly for father and mother separately whether
their first or their second computer word is used.) 
The second word for the child takes 
the other bits of the used computer words of father and mother. (Nature and
refs.2,5,6 do these steps partially in different order; we found no important
changes when we followed this computationally less efficient order.) 
The random selection of
computer words and crossing points is repeated independently for each of the
$b$ children which are born simultaneously. A mutation is active, reducing
the survival probability by a factor $1-s$, if it is dominant or if it is
carried by both computer words; otherwise the recessive mutation remains 
stored in the genome but does not affect the health.  

For the sexual selection process we go through all the females and let each
of them select randomly a male for mating. However, generalizing the methods
of Bernardes [5,6], we jump over those females who gave birth recently within
an interval of $p$ time steps; $p=2$ means that they do not mate again in the 
current time step or in the  following one. This takes into account maternal 
care for the offspring[9]. Also, males are {\it assumed} to show some degree of
marital fidelity and are not selected to mate if they have already mated before 
during the same time step. If a female cannot find a suitable male after many
attempts, she gives up for this iteration and tries again at the next time step;
otherwise she produces $b$ offspring, each of which is randomly either male or
female. Since the males have little to say in this selection process, we have
not punished them further. Thus in contrast to chapter II and to Bernardes [5,6]
we did no restrict mating to only the healthiest of them.

Fig.3 shows that for sexual reproduction with a small fraction 5/32 of dominant 
mutations there are many mutations but only few of them really count by 
diminishing the survival probability by 10 percent each; no such advantage is
seen for the corresponding asexual case. The average fitness is 0.87 in the 
sexual and only 0.81 in the asexual case; it sinks to 0.79 if all mutations
are dominant in the sexual case. Thus this bit-string Monte Carlo
simulation confirms roughly the conclusions from the simpler probabilistic 
approach of chapter II.  

\bigskip
\centerline {IV. Ageing and Sex}

In numerous species, sexual reproduction sets in only after some mimimum
reproductive age has been reached; and e.g. for women it stops after some 
maximum reproductive age (menopause). This effect can be described only by
a model incorporating age as a variable. In particular we want to explain life
after menopause.

We thus interpret the bit-string 
of the previous chapter as representing not only 32 diseases but also 
32 intervals (``years'') in the life of an individual. Thus each year we
feel the effect of at most one more hereditary disease. In contrast to 
the previous section, at a certain age we see only the bits (diseases)
corresponding to this and earlier ages; we do not yet know and feel the
diseases starting to affect us only later in life but hidden already in
our genome. 

The asexual version then becomes the well-studied Penna aging model[7] (see
ref.5 for a review), if
we replace the survival probability $(1-s)^m$ by truncation selection:
An individual survives up to the age when $T$ inherited diseases become
active; then it dies. We also use again a Verhulst factor. The sexual
reproduction follows the lines of chapter III, but again with truncation
selection and the interpretation of the bit position as the age. 

In the asexual case we know already that life stops after the maximum 
reproductive age[8] in this model, as seen most drastically for Pacific salmon.
After many generations, mutations have accumulated
in all the bits beyond the maximum age of reproduction
and then cause sudden death: The survival probability as a function of age
jumps to zero. For the sexual case, however, this no longer is true. If
the females reproduce only between the ages of 10 and 12, and the males for
all ages starting from 10 (up to certain death at age 33), then both males 
and females live until about an age of 16, roughly the same life expectancy
as for the asexual case; see fig.4a.  Fig.4b, produced on 136 processors
of the Intel Paragon at KFA J\"ulich (Germany), shows that as for the asexual
Penna model the Gompertz law[11] is reasonably confirmed: The death rate after
sexual maturity increases roughly exponentially with age.

Why can the females in this sexual model survive after their reproductive age
is passed when they cannot in the asexual model[9] ? A crucial aspect of our
model as well as nature is that sex is not transmitted genetically; independent
of the genome we take each child as male with probability 1/2, and as female
otherwise. So if death is hidden in the offspring's genes, then either both
males and females die soon, or both males and females die late. Neither nature
nor our model allows all females to die sooner from accumulated mutations than
the males. Thus, men might be useful for something (in this model).

Omitting the requirement for male fidelity does not change much, whereas 
omission of the female waiting period by setting $p=0$ causes slightly faster 
aging. (Whereas in the asexual case eventually all survivors have one common 
ancestor, this is not the case for the sexual case.) 

Separately [12] we discuss why women live longer than men, and offer somatic
(not inheritable) mutations as an explanation, in contrast to the hereditary
mutations discussed here and in most of the biological ageing literatur.  
\bigskip
\centerline {V Summary}

The Redfield algorithm is computationally fast and was easily modified to
distinguish between dominant and recessive mutations. Its exact treatment of
infinitely large populations is somewhat similar to that of Dasgupta for 
aging [13].

Various computer models have shown in this work that sexual recombination is
advantageous compared to asexual reproduction without genetic recombination.
Sex can hide recessive mutations and ensure the survival of females past
reproductive age as long as males of the same age are still able to reproduce.
Moreover, also the sexual version of the Penna model agreed with the naturally
observed Gompertz law. 
We do not claim that our models are the only ones giving this desired agreement
with nature, though we are not aware of other ``microscopic'' models explaining
the Gompertz law. ``Virtually all of many efforts to find a unifying theory of 
aging have foundered when rigorous questions were asked'' [14]. 
 
\bigskip
We thank Americo Bernardes for teaching us sex, Rosemary Redfield for explaining
her program, and CAPES, FAPERJ, FINEP, and CNPq of Brazil for support. 

\vfill \eject
\parindent=0pt
[1] A.S. Kondrashov, Nature 369, 99 (1994)

[2] R.J. Redfield, Nature 369, 145 (1994)

[3] D.Charlesworth, M.T.Morgan and B.Charlesworth, Genetic Research 61, 39
(1993)

[4] M.Lynch and W.Gabriel, Evolution 44, 1725 (1990)
 
[5] A.T.Bernardes, J.Physique I 5, 1501 (1995);
Physica A, in press; Ann.Physik (Leipzig), in press;
A.T. Bernardes, Monte Carlo Simulations of Biological Ageing, in:
{\it Annual Reviews of Computational Physics}, vol. IV, edited by D.Stauffer,
World Scientific, Singapore 1996, in press

[6] A.T.Bernardes, J. Stat. Phys., in press

[7] T.J.P.Penna, J.Stat.Phys. 78, 1629 (1995)

[8] T.J.P. Penna, S. Moss de Oliveira, and D. Stauffer, Phys.Rev. E 52, 3309 
(1995); T.J.P. Penna and S. Moss de Oliveira, J.Physique I 5, 1697
(1995)

[9] J.Thoms, P.Donahue, D.L.Hunter, and N.Jan, J.Physique I 5, 1689 (1995) 

[10] M.Walter, Staatsexamensarbeit (thesis), Cologne University (1996)

[11] M.Rose, {\it Evolutionary Biology of Aging}, Oxford University Press, 
New York 1991

[12] S. Moss de Oliveira, P.M.C. de Oliveira, and D.Stauffer, preprint for
Braz.J.Phys. 

[13] S.Dasgupta, J.Physique I 4, 1563 (1994)

[14] R.A.Lockshin and Z.Zaher, p.167 in: {\it Cellular Aging and Cell Death},
edited by N.J. Holbrook, G.R.Martin, and R.A.Lockshin, Wiley-Liss, New York 
1966.

\bigskip
Fig.1: Average survival probabilities for Redfield algorithm with dominant 
fraction $h = 0$ (diamonds), 0.2 (+), 0.4 (squares), 0.6 (x), 0.8 (triangles), 
and 1.0 (stars), for various recessive parameters r. Ref.2 corresponds to stars
at the $r=0$ axis. 
Part a gives our results for $\alpha=1$, part b for $\alpha=10$. 

Fig.2: Average survival probability for Redfield model, 
$\alpha=10$, if only males with 
less than $sextruc$ mutations (as given in figure) are selected as mates.
Part a assumes independent selection according to $0.9^m$, part b assumes
truncation selection where more than 6 mutations kill. 

Fig.3: Distribution of active genetic diseases (mutations) for the asexual and
the sexual case, $\mu = 0.3$, if 5/32 of the mutations are dominant, for the
bit-string model. Recessive mutations refer to the sexual case but do not reduce
the survival rate. 

Fig.4 Part a: Survival rate as a function of age ($T=4$); 
initially 100,000 individuals,
minimum reproduction age = 10, birth rate = 2 in asexual case and = 4 and 0 for
females and males in sexual case ($p=1,\,h=6/32$). Squares: sexual reproduction,
mutation rate = 1 per individual, maximum reproduction age = 32 for both sexes; 
triangles: same except maximum female reproduction age = 12; circles: asexual
reproduction, mutation rate = 1 per individual, maximum preproduction age = 12;
stars: same as circles except twice as high mutation rate.
Part b shows the death rate (in two definitions) logarithmically
versus age indicating the exponential increase of the Gompertz law for high
ages (6.8 million individuals of each sex initially, about 10 million between
5000 and 10000 time steps; $b=8, \, h=5/32, \, p=2, \, T=4$). 

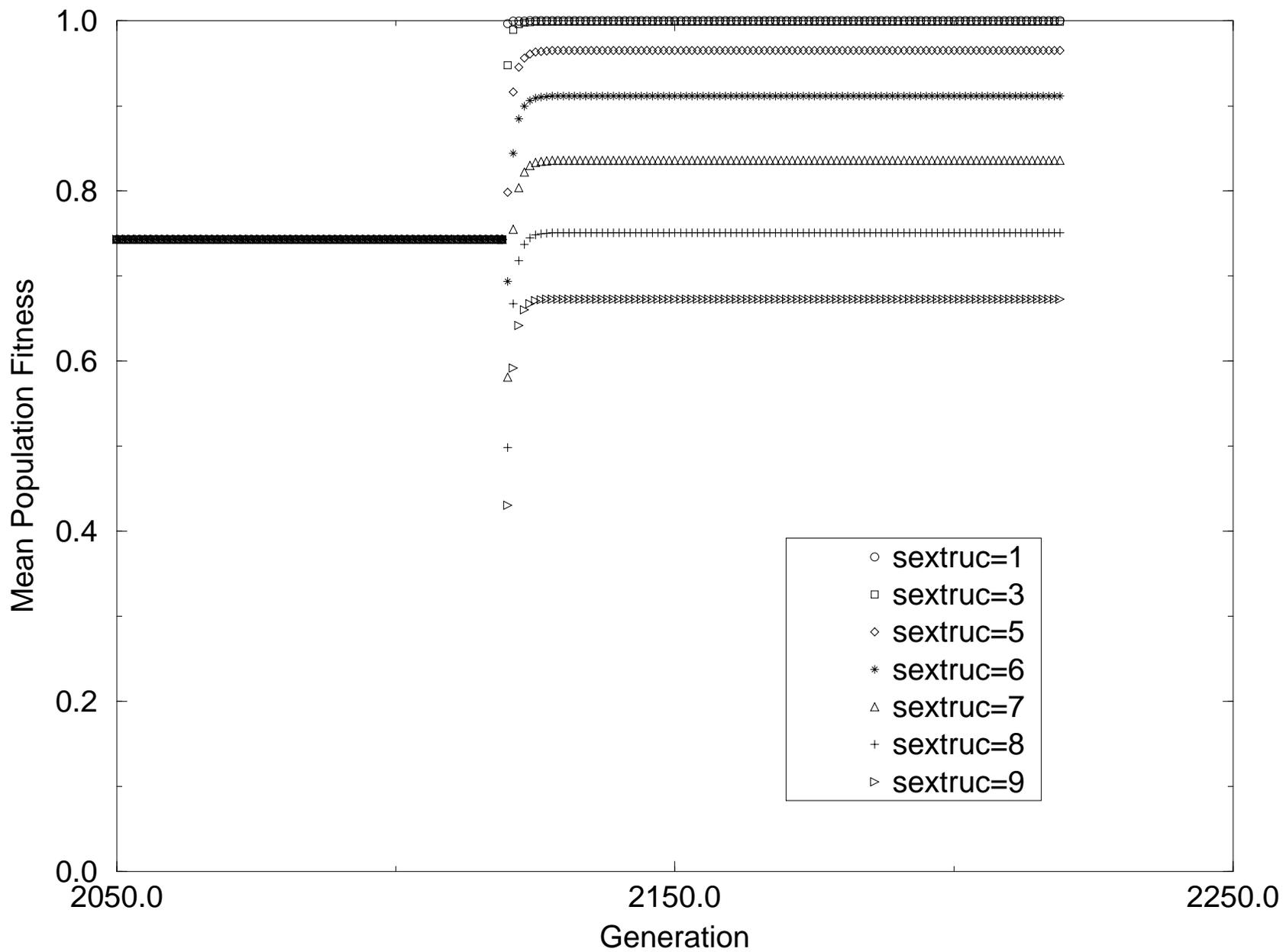
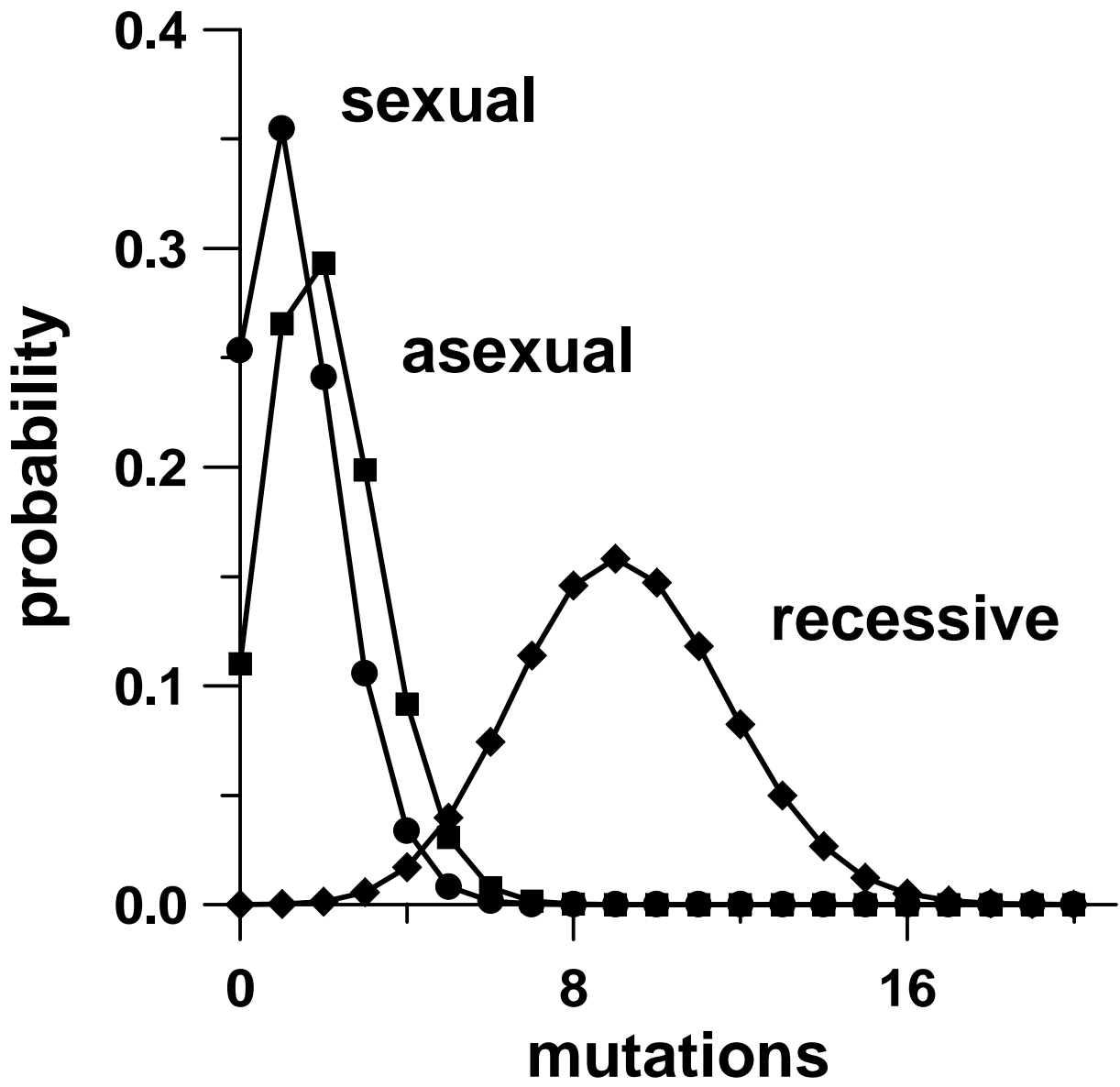
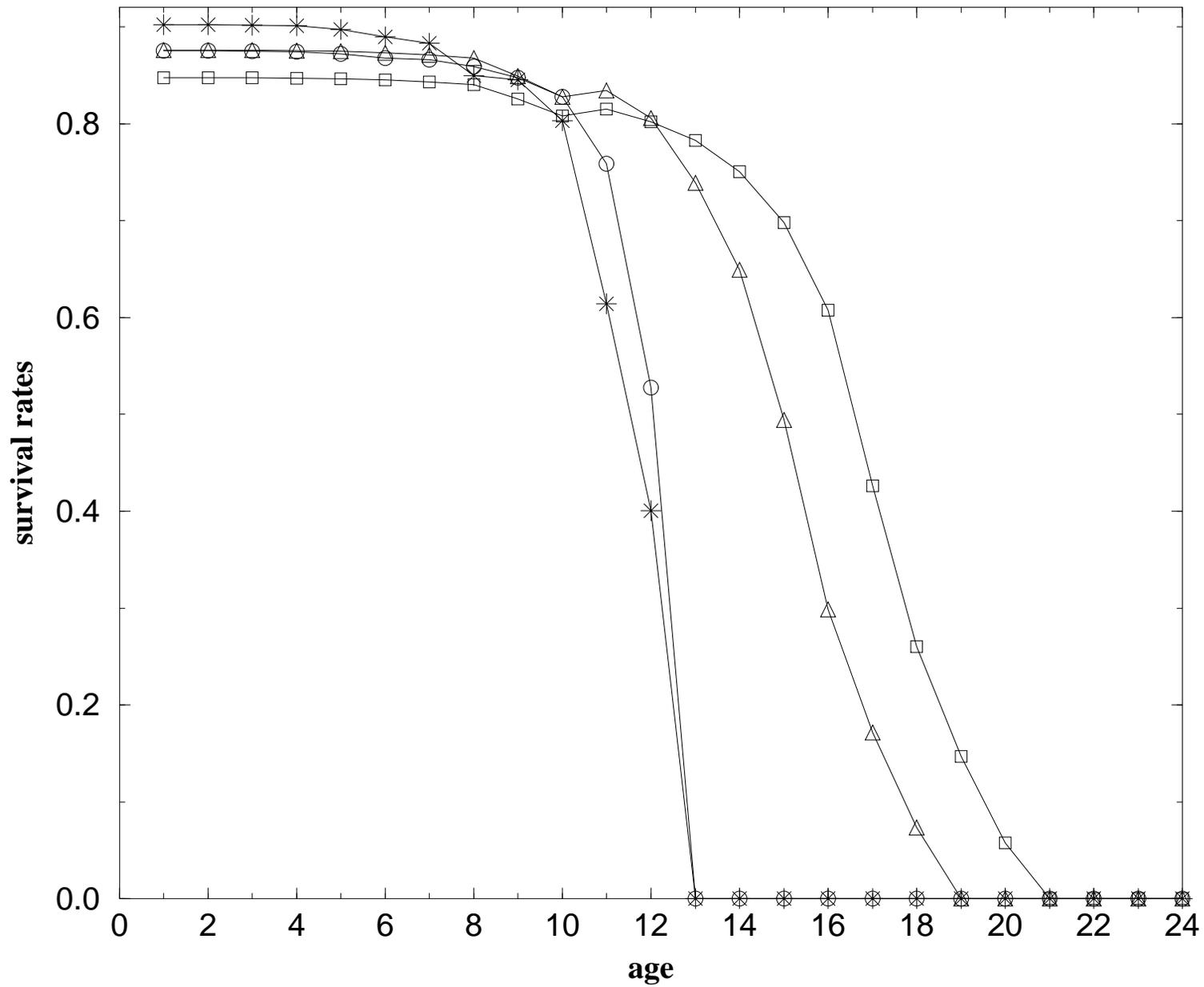
\end